# Photonics-based de-chirping and leakage cancellation for frequency-modulated continuous-wave radar system

Taixia Shi[a,b], Dingding Liang[a,b], Moxuan Han[a,b], and Yang Chen[a,b,*]

[a] Shanghai Key Laboratory of Multidimensional Information Processing, East China Normal University, Shanghai, 200241, China
[b] Engineering Center of SHMEC for Space Information and GNSS, East China Normal University, Shanghai, 200241, China
[*] ychen@ce.ecnu.edu.cn

**ABSTRACT**
A photonics-based leakage cancellation and echo signal de-chirping approach for frequency-modulated continuous-wave radar systems is proposed based on a dual-drive Mach-Zehnder modulator (DD-MZM), with its performance evaluated by the radar measurement and imaging. The de-chirp reference signal and the leakage cancellation reference signal are combined and applied to the upper arm of the DD-MZM, while the received signal including the leakage signal and echo signals is applied to the lower arm of the DD-MZM. When the amplitudes and delays of the leakage cancellation reference signal and the leakage signal are precisely matched and the DD-MZM is biased at the minimum transmission point, the leakage signal is canceled in the optical domain. The de-chirped signals are obtained after the leakage-free optical signal is detected in a photodetector. An experiment is performed. The cancellation depth of the de-chirped leakage signal is around 23 dB when the center frequency and bandwidth of the linearly frequency-modulated signal are 11.5 and 2 GHz. The leakage cancellation scheme is used in a radar system. When the leakage cancellation is not employed, the leakage signal will seriously affect the imaging results and distance measurement accuracy of the radar system. When the leakage cancellation is applied, the imaging results of multiple targets can be clearly distinguished, and the error of the distance measurement results is significantly reduced to ±10 cm.

## 1. Introduction

Frequency-modulated continuous-wave (FMCW) radar system has the advantages of the large time-bandwidth product, simple structure, low transmitting power, and strong interception resistance [1], which can be widely used in various range and speed measurement applications [2], [3]. FMCW radar is different from conventional pulse radar because it transmits and receives signals at the same time. Therefore, the FMCW radar receiver will receive a leakage signal with very strong power from the transmitter [4] besides the radar echo signals. The leakage signal is also known as a self-interference signal, whose power is much greater than that of the echo signals reflected by the targets. Thus, the leakage signal is very easy to saturate the low-noise amplifier and analog-to-digital converter in the receiver.

Since the leakage signal is much stronger, the cancellation of the leakage signal is often realized through combining multiple cancellation stages [5] from the receiving antenna to the receiver, including the antenna domain cancellation, analog domain cancellation, and digital domain cancellation. The analog domain cancellation is the key part to avoid receiver saturation. However, conventional electrical-based methods are limited in working frequency/bandwidth and lack of tunability. With the advantages of large bandwidth, high frequency, and immunity to electromagnetic interference, microwave photonics [6] is a promising technique for processing microwave signals [7].

In addition to the FMCW radar system, the in-band full-duplex (IBFD) communication system also suffers from the strong leakage signal due to simultaneous transmitting and receiving signals at the same frequency band [8]. The analog domain cancellation in both IBFD communication systems and FMCW radar systems is mainly based on constructing reference signals from the known transmitted signals to cancel the leakage signals. There are many similarities between IBFD communication systems and FMCW radar systems in the leakage cancellation studies. In the past few years, many studies [9], [10] have been reported for different key problems in the photonics-based self-interference cancellation (SIC) system for the cancellation of the leakage signal in IBFD communication systems. In [11]–[13], the self-interference can be canceled by using two parallel Mach–Zehnder modulators (MZMs)/equivalent intensity modulators biased at opposite quadrature transmission points. The SIC can also be achieved by using two modulators or modulated lasers and a balanced photodetector [14]–[17]. To avoid the instability of the SIC system with two optical paths, SIC methods [18]–[25] based on a single modulator and a photodetector are proposed. Furthermore, SIC methods combining frequency downconversion are also widely studied [22]–[25]. However, all these methods mentioned above are not specially designed for the FMCW radar system. Only Ref. [26] studied the leakage cancellation for the FMCW radar system using a polarization-division-multiplexed MZM, and a 17.5-dB leakage cancellation depth at 15 GHz with a bandwidth of 2 GHz was achieved. Although the leakage cancellation method in [26] is designed for the FMCW radar system, there is no significant difference from the leakage cancellation method which was previously applied to IBFD communication systems. The unique characteristics of the FMCW radar are still not reflected in [26], such as the de-chirping of the echo signals and the effect of leakage cancellation on the imaging and distance measurement of the FMCW radar systems. Thus, it is highly desirable that the cancellation performance of the leakage cancellation system is verified by taking the radar application, such as distance measurement and imaging, into consideration.

Furthermore, in FMCW radar systems, de-chirping processing is required for further radar signal processing, whereas the original spectrum of the echo signal does not to be fully recovered. In fact, because the echo signals reflected from the targets commonly travel a much longer distance than the leakage signal between the two antennas, the frequency component corresponding to the echo signals and that corresponding to the leakage signal is separated in the de-chirped signals after the de-chirping process. When the de-chirping function is incorporated in the leakage cancellation system, the cancellation in the analog domain is mainly to prevent the receiver from being saturated besides reducing the power of the de-chirped leakage signal, which is much different from the leakage cancellation in the IBFD communication systems. In the IBFD communication systems, no de-chirping process is needed. Therefore, the original spectrum of the signal of interest has to be separated from that of the leakage signal directly. Therefore, it is very worthwhile to study the radar leakage cancellation in conjunction with the radar de-chirping.

In this paper, a photonics-based leakage cancellation and echo signals de-chirping method for FMCW radar systems is proposed. The leakage cancellation structure of this approach is simple. The leakage is directly canceled in the optical domain in a dual-drive Mach-Zehnder modulator (DD-MZM) and the echo signals are de-chirped after photodetection. To the best of our knowledge, this is the first photonics-based radar leakage cancellation and echo signals de-chirping system, and the effect of the leakage cancellation on the imaging and distance measurement of the FMCW radar systems is studied for the first time. The power of the de-chirp reference signal is adjusted to decrease the de-chirped signal background noise, so the cancellation performance is optimized. After the optimization, the cancellation depth of the leakage in the obtained de-chirped signal is around 23 dB when the center frequency and bandwidth of the linearly frequency-modulated (LFM) signal are 11.5 and 2 GHz. The leakage cancellation scheme is used in a radar system. When the leakage cancellation is not employed, the leakage signal will seriously affect the imaging results and distance measurement accuracy of the radar system. When the leakage cancellation is applied, the imaging results of multiple targets can be clearly distinguished, and the error of the distance measurement results is significantly reduced to ±10 cm.

## 2. Principle

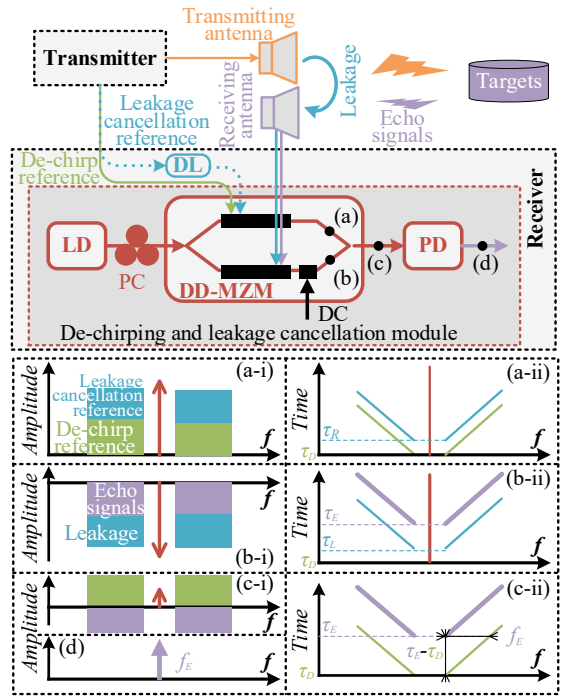

Fig. 1. Schematic diagram of the proposed leakage cancellation and de-chirping system. LD, laser diode; PC, polarization controller; DD-MZM, dual-drive Mach-Zehnder modulator; DL, delay line; DC, direct current; PD, photodetector. (a)-(d) are the schematic diagrams of the signals at different locations in the system diagram.

The schematic diagram of the leakage cancellation and de-chirping method for the FMCW radar system is shown in Fig. 1. An optical signal generated from a continuous-wave (CW) laser diode (LD) is sent to a DD-MZM via a polarization controller (PC). The de-chirp reference signal and the leakage cancellation reference signal tapped from the radar transmitter are sent to the upper arm of the DD-MZM, whereas the received signals from the receiving antenna, including the leakage signal and the radar echo signals, are sent to the lower arm of the DD-MZM. The DD-MZM is biased at the minimum transmission point (MITP) to cancel the leakage signal in the optical domain. The leakage-free optical signal from the DD-MZM is injected into a photodetector (PD), where the echo signals and the de-chirp reference signal are beaten and the de-chirped signal for radar signal processing is generated. By using such a simple system, the leakage cancellation and radar signal de-chirping are simultaneously implemented.

It is assumed that the start angular frequency and chirp rate of the LFM signal are $\omega_s$ and $k$. In this case, the de-chirp reference signal, the leakage cancellation reference signal, the leakage signal, and the echo signal can be expressed as

$$V_D(t) = A_D \cos\left[\omega_s(t-\tau_D) + \pi k(t-\tau_D)^2\right], \tag{1}$$

$$V_R(t) = A_R \cos\left[\omega_s(t-\tau_R) + \pi k(t-\tau_R)^2\right], \tag{2}$$

$$V_L(t) = A_L \cos\left[\omega_s(t-\tau_L) + \pi k(t-\tau_L)^2\right], \tag{3}$$

$$V_E(t) = A_E \cos\left[\omega_s(t-\tau_E) + \pi k(t-\tau_E)^2\right], \tag{4}$$

where $A_D$, $A_R$, $A_L$, and $A_E$ are the amplitudes, $\tau_D$, $\tau_R$, $\tau_L$, and $\tau_E$ are the delays of the corresponding signals. To simplify the following analysis, it is assumed that

$$\theta_D = \omega_s(t-\tau_D) + \pi k(t-\tau_D)^2, \tag{5}$$

$$\theta_R = \omega_s(t-\tau_R) + \pi k(t-\tau_R)^2, \tag{6}$$

$$\theta_L = \omega_s(t-\tau_L) + \pi k(t-\tau_L)^2, \tag{7}$$

$$\theta_E = \omega_s(t-\tau_E) + \pi k(t-\tau_E)^2. \tag{8}$$

Therefore, the de-chirp reference signal, the leakage cancellation reference signal, the leakage signal, and the echo signal can be expressed as $A_D\cos(\theta_D)$, $A_R\cos(\theta_R)$, $A_L\cos(\theta_L)$, and $A_E\cos(\theta_E)$, respectively. The input optical signal from the LD is assumed to be $\exp(j\omega_c t)$, so the optical signal from the upper arm of the DD-MZM can be expressed as

$$E_{upper}(t) = \frac{1}{\sqrt{2}} \exp[j\omega_c t + jm_D \cos(\theta_D) + jm_R \cos(\theta_R)]$$
$$\approx \frac{1}{\sqrt{2}} \exp(j\omega_c t)[J_0(m_D)J_0(m_R)$$
$$+ 2jJ_1(m_D)J_0(m_R)\cos(\theta_D) \qquad (9)$$
$$+ 2jJ_0(m_D)J_1(m_R)\cos(\theta_R)$$
$$- 4J_1(m_D)J_1(m_R)\cos(\theta_D)\cos(\theta_R)],$$

where $J_n(\cdot)$ is the $n$th-order Bessel function of the first kind, $m_D = \pi A_D/V_\pi$ and $m_R = \pi A_R/V_\pi$ are the modulation indices, and $V_\pi$ represents the half-wave voltage of the DD-MZM. In the derivation of Eq. (9), small-signal modulation condition ($m_D \ll 1$, $m_R \ll 1$) is applied, so only the first-order optical sidebands are taken into account. The spectrum and time-frequency diagram of the optical signal from the upper arm of the DD-MZM are shown in Fig. 1(a). When the DD-MZM is biased at the MITP, the optical signal from the lower arm of the DD-MZM is given by

$$E_{lower}(t) = \frac{1}{\sqrt{2}} \exp[j\omega_c t + jm_L \cos(\theta_L) + jm_E \cos(\theta_E) + \pi]$$
$$\approx -\frac{1}{\sqrt{2}} \exp(j\omega_c t)[J_0(m_L)J_0(m_E)$$
$$+ 2jJ_1(m_L)J_0(m_E)\cos(\theta_L) \qquad (10)$$
$$+ 2jJ_0(m_L)J_1(m_E)\cos(\theta_E)$$
$$- 4J_1(m_L)J_1(m_E)\cos(\theta_L)\cos(\theta_E)],$$

where $m_L = \pi A_L/V_\pi$ and $m_E = \pi A_E/V_\pi$ are the modulation indices. In the derivation of Eq. (10), small-signal modulation condition ($m_L \ll 1$, $m_E \ll 1$) is applied. The spectrum and time-frequency diagram of the optical signal from the lower arm of the DD-MZM are shown in Fig. 1(b). Considering the small-signal modulation conditions used before, we have $J_0(m_D) \approx 1$, $J_0(m_R) \approx 1$, $J_0(m_L) \approx 1$, $J_0(m_E) \approx 1$, $J_1(m_D) \ll J_0(m_D)$, $J_1(m_R) \ll J_0(m_R)$, $J_1(m_L) \ll J_0(m_L)$, $J_1(m_E) \ll J_0(m_E)$, so the optical signal from the DD-MZM can be simplified as

$$E_{DD\text{-}MZM}(t) \approx \exp(j\omega_c t)\{jJ_1(m_D)\cos(\theta_D) + jJ_1(m_R)\cos(\theta_R)$$
$$- jJ_1(m_L)\cos(\theta_L) - jJ_1(m_E)\cos(\theta_E)\}. \qquad (11)$$

When $\tau_R = \tau_L$ and $A_R = A_L$ establish, the leakage signal is canceled by the leakage cancellation reference signal in the optical domain, and the optical signal from the DD-MZM can be expressed as

$$E_{DD\text{-}MZM}(t) = \exp(j\omega_c t)[jJ_1(m_D)\cos(\theta_D) - jJ_1(m_E)\cos(\theta_E)]. \qquad (12)$$

As can be seen, besides the leakage signal, the optical carrier from the two arms of the DD-MZM is also canceled. The spectrum and time-frequency diagram corresponding to the optical signal from the DD-MZM are shown in Fig. 1(c). Then, the leakage-free optical signal from the DD-MZM is detected in the PD with a responsivity of $\Re$. The photocurrent from the PD can be written as

$$\begin{aligned}
i_{PD}(t) &= \Re\left|E_{DD\text{-}MZM}(t)\right|^2 \\
&= \frac{\Re}{2}\{J_1^2(m_D)[1+\cos(2\theta_D)] + J_1^2(m_E)[1+\cos(2\theta_E)] \\
&\quad -2J_1(m_D)J_1(m_E)[\cos(\theta_E+\theta_D)+\cos(\theta_D-\theta_E)]\}.
\end{aligned} \quad (13)$$

After low-pass filtering and assuming that $\tau_D = 0$, the photocurrent from the PD can be expressed as

$$\begin{aligned}
i_{LPF}(t) &= \frac{\Re}{2}\left[J_1^2(m_D)+J_1^2(m_E)-2J_1(m_D)J_1(m_E)\cos(\theta_D-\theta_E)\right] \\
&= \frac{\Re}{2}J_1^2(m_D)+\frac{\Re}{2}J_1^2(m_E) \\
&\quad -\Re J_1(m_D)J_1(m_E)\cos\left[\omega_s\tau_E + 2\pi kt\tau_E - \pi k\tau_E^2\right].
\end{aligned} \quad (14)$$

The de-chirped signal at the low-frequency band is obtained, which can be expressed as

$$f_E = k\tau_E. \quad (15)$$

The spectrum corresponding to the de-chirped electrical signal from the PD is shown in Fig. 1(d).

For the FMCW radar system, the distance measurement and the ISAR imaging can be implemented by further processing the de-chirped signals. According to Eq. (11), the distance can be derived as

$$R = \frac{1}{2}c\tau_E = \frac{c}{2k}f_E, \quad (16)$$

where $c$ is the velocity of light in a vacuum. To implement ISAR imaging, when the radar transmits a sequence of N periods and each period of the de-chirped signal has M samples, the two-dimension image can be constructed by the rearranged M×N matrix from the N·M samples [27], [28]. The ISAR imaging range resolution and the cross-range can be expressed as

$$R_R = \frac{c}{2B}, \quad (17)$$

$$R_c = \frac{\lambda}{2T_i\Omega}, \quad (18)$$

where $B$, $\lambda$, $T_i$, and $\Omega$ are the LFM signal bandwidth, the LFM signal center wavelength, the integration time, and the rotating speed of the imaging targets, respectively. In the following of this paper, the leakage cancellation performance of the system is studied not only in the spectrum as common in the previously reported works, but also in the results of radar detection and ranging.

## 3. Experiment and results

3.1 Experimental setup

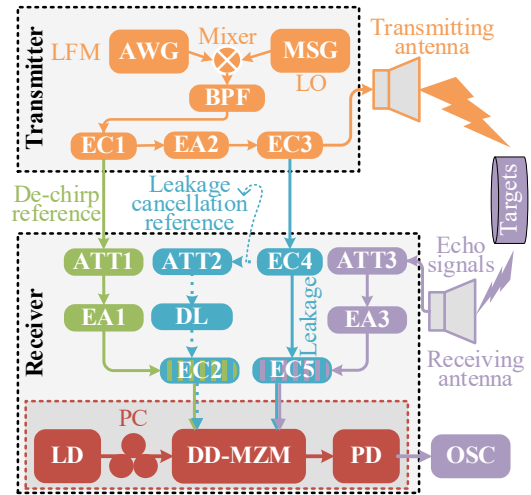

Fig. 2. Experimental setup of the proposed leakage cancellation and de-chirping system. AWG, arbitrary waveform generator; MSG, microwave signal generator; LFM, linearly frequency-modulated signal; LO, local oscillator signal; BPF, bandpass filter; EC, electrical coupler; EA, electrical amplifier; ATT, attenuator; DL, delay line; LD, laser diode; PC, polarization controller; DD-MZM, dual-drive Mach-Zehnder modulator; PD, photodetector; OSC, oscilloscope.

An experiment is carried out based on the setup shown in Fig. 2. A 15.5-dBm CW light wave generated from an LD (ID Photonics CoBriteDX1-1-C-H01-FA) with a wavelength of 1548.9 nm is injected into a DD-MZM (Fujitsu FTM7937EZ200) via a PC. All the LFM signals including the leakage signal, the transmitted signal, the de-chirp reference signal, and the leakage cancellation reference signal are generated from an arbitrary waveform generator (AWG, Keysight M8190A), whereas the local oscillator (LO) signal is generated from a microwave signal generator (MSG, Agilent 83630B).

An intermediate frequency (IF) LFM signal from the AWG is firstly up-converted to the higher frequency band using a 7-dBm LO signal via an electrical mixer (M/A-COM M14A) and then the up-converted LFM signal is filtered by an electrical bandpass filter (BPF, KGL YA356-2, 10.4-14.1 GHz). The signal from the BPF is divided into two parts by a 3-dB electrical coupler (EC1, MCLI PS2-11, 2-18 GHz, -3 dB). One output from EC1 is sent to EC2 (Narda 4456-2, 2-18 GHz, -3 dB) through a tunable electrical attenuator (ATT1, Norsal IND 7131-10, 10-20 GHz) and an electrical amplifier (EA1, ALM 145-5023-293 5.85-14.5 GHz) as the de-chirp reference signal. The other output from EC1 is split into two parts by a directional coupler (EC3, KRYTAR, MODEL 1818, 2-18 GHz, -16 dB) via EA2 (CLM 145-7039-293B, 5.85-14.50 GHz). The coupled output of EC3 is split into two parts again by EC4 (AEROFLEX YL-56, 7-14GHz, -3 dB). One output from EC4 is sent to EC2 through ATT2 (Narda 4799, 4-18GHz) and a tunable electrical delay line (DL, Sage 6705) as the leakage cancellation reference signal. The de-chirp reference signal and the leakage cancellation reference signal are combined at EC2 and then applied to the upper arm of the DD-MZM.

Another output of EC4 is injected into EC5 (Narda 4456-2, 2-18 GHz, -3 dB) to simulate a stronger leakage signal whose power is much greater than that of the echo signals. The direct output of EC3 is fed to a transmitting antenna (GHA080180-SMF-14, 8-18 GHz) and then radiated for targets detection. The echo signals reflected by the targets are collected by a receiving antenna (GHA080180-SMF-14, 8-18 GHz). The echo signals are passing through ATT3 (Norsal IND 7131-10, 10-20 GHz) and EA3 (ALM 145-5023-293 5.85-14.5 GHz) and then combined with the strong leakage signal at EC5 to simulate the received signals, which is then applied to the lower arm of the DD-MZM.

It should be noted that ATT1 and EA1 are used to adjust the power of the de-chirp reference signal for system performance optimization; ATT3 and EA3 are used to adjust the power of the echo signals to study the system performance; ATT2 and the DL are used to match the amplitudes and delays of the leakage cancellation reference signal and leakage signal for leakage cancellation. When the amplitudes and delays of the leakage cancellation reference signal and leakage signal are precisely matched and the DD-MZM is biased at MITP, the leakage signal can be canceled in the optical domain. Then, the optical signal from the DD-MZM is injected into a PD (Nortel Networks PP-10G). The waveforms of the de-chirped signal from the PD are captured by an OSC (R&S RTO2032) by implementing a digital low-pass filter in the OSC. The target range and imaging results are extracted from the waveforms of the de-chirped signals.

3.2 Cancellation performance without echo signals

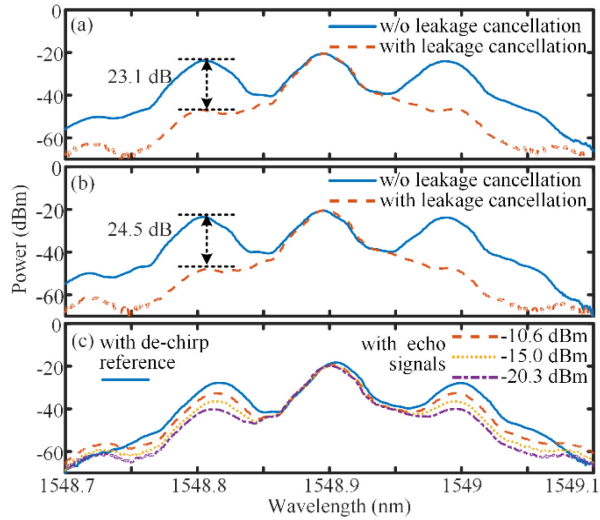

Fig. 3. Optical spectra at the output of the DD-MZM with and without leakage cancellation when the bandwidth of the LFM signal are (a) 2 GHz, (b) 1 GHz. (c) Optical spectra at the output of the DD-MZM with different de-chirp reference or echo signal power.

Firstly, the leakage cancellation performance without echo signals is investigated both in the optical domain and the electrical domain. It should be pointed out here that for the convenience of description and experiment, all subsequent echo signal power in this paper refers to the power of the echo signal input to EC5 after the echo signal is received by the receiving antenna, attenuated by ATT3, and amplified by EA3.

To investigate the leakage cancellation performance in the optical domain, the amplitude of the IF LFM signal from the AWG is set to 500 mV, the period of the IF LFM signal is set to 0.1 ms, the center frequency of the IF LFM signal is set to 2 GHz, the bandwidth of the IF LFM signal from is set to 1 or 2 GHz, and the frequency of the LO signal from the MSG is set to 9.5 GHz. In this case, the center frequency of the LFM signal from the BPF is 11.5 GHz, the bandwidth of the LFM signal from the BPF is 1 or 2 GHz. The power of the leakage signal is around -3 dBm. The optical signal from the DD-MZM is observed using an optical spectrum analyzer (OSA, ANDO AQ6317B). Fig. 3(a) shows the optical spectra of the optical signal from the DD-MZM with and without leakage cancellation when the de-chirp reference signal and the echo signals are not applied to the DD-MZM. The blue solid line in Fig. 3(a) shows the optical spectrum without leakage cancellation by disconnecting the leakage cancellation reference signal from the system, whereas the red dashed line in Fig. 3(a) shows the optical spectrum with leakage cancellation when the bandwidth of the LFM signal is 2 GHz. It can be seen from the optical domain that the cancelation depth of the leakage sideband is around 23.1 dB. It should be noted that the optical carrier is suppressed since the DD-MZM is biased at MITP. In Fig. 3, the carrier and the sidebands have close power because the power of the leakage signal is as low as -3 dBm. When the bandwidth of the LFM signal is decreased to 1 GHz, the cancelation depth of the leakage sideband changes to 24.5 dB, as shown in Fig 3(b). The blue solid line in Fig. 3(c) shows the optical spectrum when only the de-chirp reference signal with a power of -5.3 dBm is applied to the DD-MZM. The red dashed line, yellow dotted line, and purple dash-dotted line in Fig. 3(c) show the optical spectra when only the echo signals with a power of -10.6, -15.1, and -20.3 dBm is applied to the DD-MZM. It should be noted that the echo signals received by the receiving antenna are amplified by using EA3 and the power of the echo signals is adjusted in the experiment by using ATT3 following the receiving antenna.

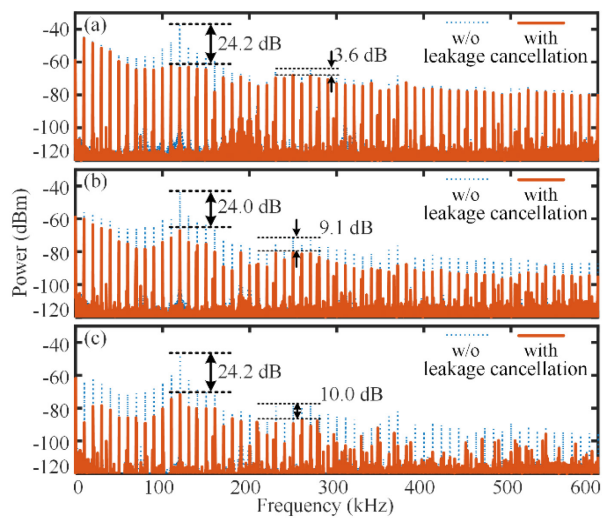

Fig. 4. Electrical spectra of the de-chirped signal with and without leakage cancellation when the power of the de-chirp reference signal is (a) -0.8 dBm (b) -5.3 dBm (c) -9.6 dBm.

The de-chirped signals can be obtained after photodetection when the de-chirp reference signal is applied. When the de-chirp reference signal is enabled, the leakage cancellation performance with different de-chirp reference signal power is investigated. In this study, the echo signals are also not applied. The de-chirp reference signal power is adjusted by tuning ATT1. The center frequency and bandwidth of the LFM signal from the BPF are 11.5 and 2 GHz, respectively. The power of the leakage signal is -3 dBm. The waveforms of the de-chirped signals from the PD are captured by the OSC with a 4-MSa/s sampling rate, and the electrical spectra of the de-chirped signals are obtained from the waveforms using the fast Fourier transform. The electrical spectra of the de-chirped signal with and without leakage cancellation are shown in Fig. 4. Fig. 4(a) shows that the leakage and de-chirped signal background noise cancellation depths are 24.2 and 3.6 dB when the power of the de-chirped signal is -0.8 dBm. When the power of the de-chirped signal is decreased to -5.3 dBm, the de-chirped signal background noise cancellation depth is increased to 9.1 dB and there is no significant change in the leakage cancellation depth, as shown in Fig. 4(b). When the power of the de-chirped signal further is decreased to -9.6 dBm, the de-chirped signal background noise cancellation depth is increased to 10 dB while the leakage cancellation depth has still no significant change, as shown in Fig. 4(c). It is indicated from Fig. 4(a)-(c) that the power of the de-chirped signal almost does not affect the depth of leakage cancellation. Smaller de-chirp reference signal power introduces less de-chirped signal background noise and also smaller power of the de-chirped signal. Therefore, the de-chirp reference signal power should be properly set. According to these results, the system performance optimization can be achieved by adjusting the de-chirp reference signal power. If the power of the de-chirp reference signal is not properly set, the leakage cancellation with a larger leakage cancellation depth does not guarantee improved imaging and radar measurement performance because of the de-chirped signal background noise.

In Fig. 4, the frequency component in the de-chirped signal corresponding to the leakage signal is at around 120 kHz, which corresponds to a distance of 1.8 m. These two values are indeed determined by the cable lengths used in the experiment.

3.3 Leakage cancellation and ISAR imaging

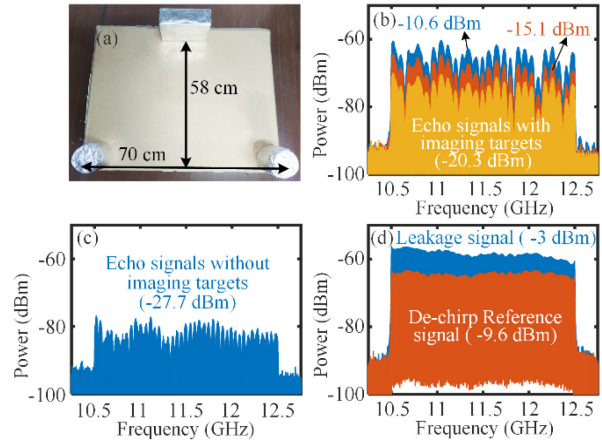

Fig. 5. (a) One cuboid and two cylinders packed with silver papers. Electrical spectra of the echo signals (b) with and (c) without imaging targets. (d) Electrical spectra of the leakage signal and the de-chirp reference signal.

Then, the imaging targets are employed in the experiment. The leakage cancellation performance with the echo signals reflected by the imaging targets and the effect of leakage cancellation on ISAR imaging are further studied. One cuboid and two cylinders packed with silver paper are used as the targets, and the imaging targets are placed on the turntable, as shown in Fig. 5(a). The distance between the center of the turntable and the antenna pair is around 1.89 m.

The electrical spectra and power of the LFM signals applied to the DD-MZM are measured using an electrical spectrum analyzer (ESA, R&S FSP-40). When measuring the electrical spectra and power of the LFM signals, the turntable is not rotated. The center frequency and bandwidth of the LFM signal from the BPF are 11.5 and 2 GHz, respectively. The power of the leakage signal is -3 dBm. The power of the de-chirp reference signal is set to -9.6 dBm. The power of the LFM signal fed into the transmitting antenna is around 20 dBm, while that of the echo signals received from the receiving antenna is adjusted by tuning ATT3 to investigate the system performance with different echo signal power.

The electrical spectra of the echo signals with different power of -10.6, -15.1, and -20.3 dBm are shown in Fig. 5(b). Many ripples are observed in the spectra, which are mainly caused by the superposition of echo signals from different reflection paths. The ripples are similar in shape for different power of the echo signals because the position of the reflecting objects does not change when the power is adjusted. Fig. 5(c) shows the electrical spectra of the echo signals when the targets are removed from the turntable. It can be seen that the electrical spectrum shown in Fig. 5(c) has more ripples when the imaging targets are removed, which are caused by the reflection of the LFM signals in the background environment. The blue line in Fig. 5(d) shows the electrical spectrum of a -3-dBm leakage signal, whereas the red line in Fig. 5(d) shows the electrical spectrum of a -9.6-dBm de-chirp reference signal. As can be seen, the two spectra shown in Fig. 5(d) have very similar spectrum shapes.

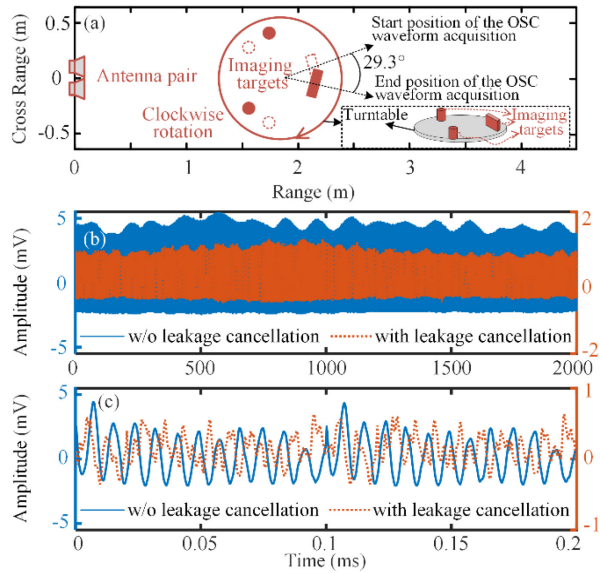

Fig. 6. (a) The schematic diagram of the targets imaging. (b) The temporal waveform of the de-chirped signal with and without leakage cancellation in 2000 ms. (c) A section temporal waveform of (a) from 0 to 0.2 ms.

To investigate the leakage cancellation performance and the effect of leakage cancellation on ISAR imaging, the imaging targets rotate clockwise with the turntable with a period of 24.56 s. The de-chirped signal is captured by the OSC with a 4-MSa/s sampling rate, and the sampling time of the de-chirped signals is 2000 ms. Fig. 6(a) shows the schematic diagram of the targets imaging. Since the OSC sampling time is 2000 ms, the turntable rotates around 29.3 degrees. The center frequency and bandwidth of the LFM signal from the BPF are 11.5 and 2 GHz, respectively. The period of the LFM signal is set to 0.1 ms. The power of the leakage signal is -3 dBm. The power of the de-chirp reference signal is set to -9.6 dBm. Fig. 6(b) shows the temporal waveform of the de-chirped signal with and without leakage cancellation when the echo signal power is -20.3 dBm. After the leakage cancellation, the waveform amplitude becomes significantly smaller. Furthermore, the waveform amplitude changes with time, and the waveform amplitude reaches the maximum at about 1000 ms after leakage cancellation, which is because the cuboid is directly in front of the antenna and can reflect more echo signals at this time. Fig. 6(c) shows a section of the temporal waveform from 0 to 0.2 ms in two periods. The waveforms are very similar at two adjacent periods due to the small rotation angle of the turntable within 0.2 ms.

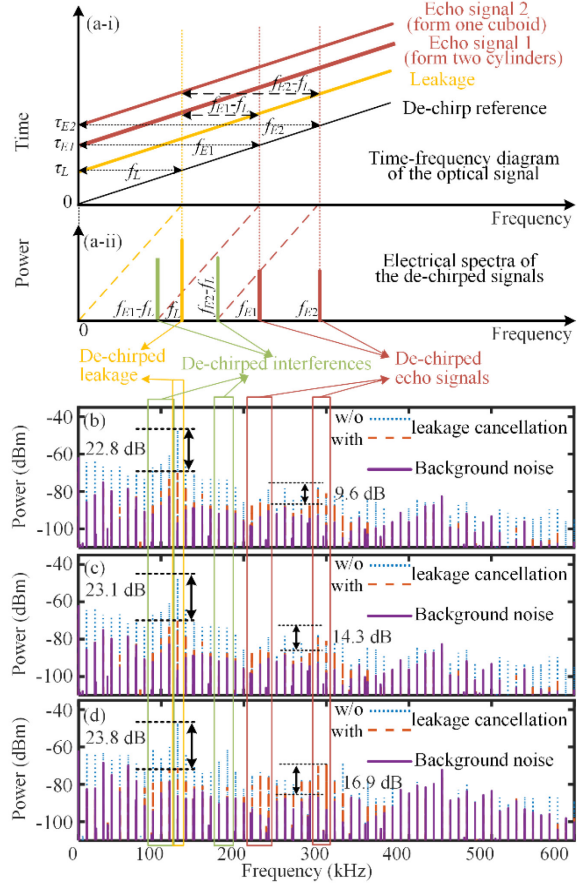

Fig. 7. (a) Schematic diagram of the photonic de-chirping process without leakage cancellation. Electrical spectra of the de-chirped signal with and without leakage cancellation when the power of the echo signal is (b) -20.3 dBm, (c) -15.1 dBm, (d) -10.6 dBm.

Fig. 7(a) shows the schematic diagram of the photonic de-chirping process when the leakage cancellation is not employed. Fig. 7(a-i) shows the time-frequency diagram of the +first-order optical sidebands of the optical signal from the DD-MZM when the cuboid is furthest away from the antenna pair and the two cylinders are closest to the antenna pair. The corresponding diagram of the -first-order sideband is similar and not shown here. Fig. 7(a-ii) shows the electrical spectra of the de-chirped signals. It can be seen that besides the de-chirped leakage signal and the de-chirped echo signals, the de-chirped interferences are introduced owning to the beating between the first-order optical sideband of the leakage and the first-order sideband of the echo signals. Thus, the de-chirped signals mainly include the de-chirped leakage signal, the de-chirped echo signals, and the de-chirped interferences.

The electrical spectra of the de-chirped signals can be obtained from the waveforms using the fast Fourier transform. The leakage cancellation performance with different echo signal power is studied in the electrical domain. The echo signal power is adjusted by tuning ATT3. The center frequency and bandwidth of the LFM signal from the BPF are 11.5 and 2 GHz, respectively. The power of the leakage signal is -3 dBm. The power of the de-chirp reference signal is set to -9.6 dBm. Fig. 7(b) shows the corresponding electrical spectrum of Fig. 6(b) when the echo signal power is -20.3 dBm. The blue dotted line in Fig. 7(b) shows the electrical spectrum without leakage cancellation, and the frequency of the de-chirped leakage signal is around 120 kHz. The frequency range of the de-chirped echo signal corresponding to the cuboid is around from 290 to 300 kHz, and the frequency range of the de-chirped echo signals corresponding to the two cylinders is around from 210 to 230 kHz. Since the distance between the antenna pair and the three imaging targets varies, as shown in Fig. 6(a), the de-chirped echo signals frequency corresponds to a frequency range. Furthermore, the frequency ranges of the de-chirped interferences are around from 90 to 110 kHz and from 170 to 180 kHz. It can be seen that the de-chirped interferences also influence the acquisition of the echo signal information when the leakage signal is not canceled. In the experiment, the setup, especially the cable lengths are carefully selected to make these three mentioned signals not overlap with each other. In real-world FMCW radar systems, the three signals are naturally separated because the echo signal always travels a much longer distance than the leakage signal.

The purple solid line in Fig. 7(b) shows the electrical spectrum of the de-chirped signal background noise when the leakage signal is not applied to the DD-MZM and the imaging targets are removed. The red dashed line in Fig. 7(b) shows the electrical spectrum with leakage cancellation. As can be seen, the leakage cancellation depth is around 22.8 dB, and the de-chirped interference is also much suppressed. The de-chirped interferences are decreased because the leakage is canceled in the optical domain and the de-chirped interference is obtained by the beating between the first-order sideband of the leakage and the echo signals. As can be seen, the de-chirped signal of the targets is around 9.6 dB higher than the de-chirped signal background noise. When the echo signal power is increased to -15.1 dBm, the cancellation depth is increased slightly to 23.1 dB and the de-chirped signal of the targets is 14.3 dB higher than the de-chirped signal background noise, as shown in Fig. 7(c). When the echo signal power is further increased to -10.6 dBm, the cancellation depth increases to 23.8 dB and the de-chirped signal of the targets is 16.9 dB higher than the de-chirped signal background noise, as shown in Fig. 7(d). It can be summarized that the de-chirped leakage signal and the de-chirped interferences can be effectively suppressed after the leakage cancellation.

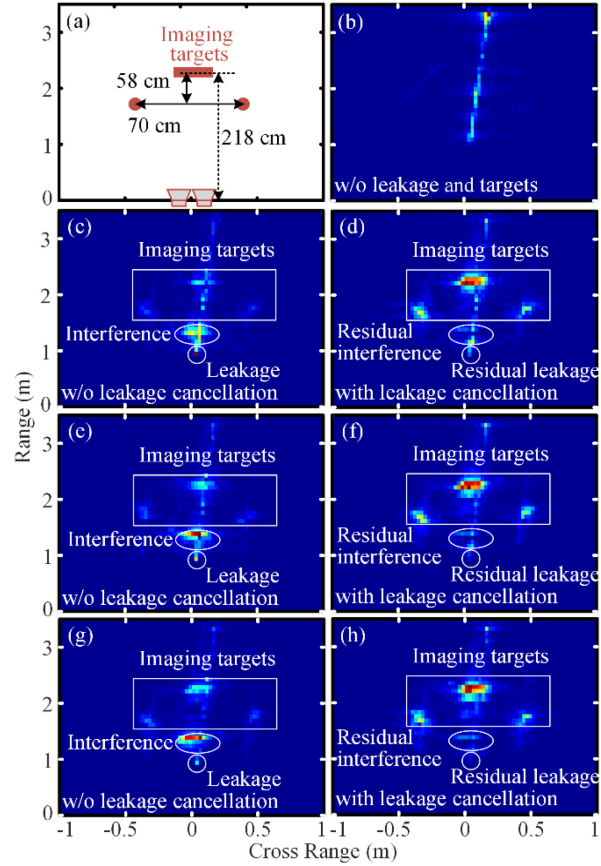

Fig. 8. (a) Schematic diagram of the positions of the three imaging targets. (b) The imaging result when the leakage signal and three imaging targets are not employed. The imaging results of the three targets (c) without and (d) with leakage cancellation when the echo signal power is -20.3 dBm, (e) without and (f) with leakage cancellation when the echo signal power is -15.1 dBm, (g) without and (h) with leakage cancellation when the echo signal power is -10.6 dBm.

Afterward, the leakage cancellation performance is evaluated by implementing the ISAR imaging. The waveforms of the de-chirped signals captured by the OSC are post-processed using Matlab, and a digital high-pass filter with a 140-kHz 3-dB cutoff frequency is also used to further suppress the residual de-chirped leakage signal and de-chirped interferences. According to (17) and (18), The theoretical range resolution and cross-range resolutions are 7.5 cm and 2.55 cm, respectively. The center frequency and bandwidth of the LFM signal from the BPF are 11.5 and 2 GHz, respectively. The power of the leakage signal is -3 dBm. The power of the de-chirp reference signal is set to -9.6 dBm.

Fig. 8(a) shows the schematic diagram of the positions of the three imaging targets. Fig. 8(b) shows the imaging result when the leakage signal is not applied and the imaging targets are removed, which is used to show the influence of the background environment on the imaging results. The imaging results of the three targets without and with leakage cancellation are shown in Fig. 8(c) and (d) when the echo signal power is set to -20.3 dBm by amplification using EA3 and attenuation using ATT3. It can be seen that the imaging results are very blurred when the leakage cancellation is not applied since the leakage and the de-chirped interference are too strong. It should be noted that the strength of the interference is depending on the leakage. When the leakage is canceled in the optical domain, the influence of de-chirped interference signal on imaging is also greatly weakened. The interference shown in Fig. 8(c) is corresponding to the cuboid. The interference corresponding to the two cylinders is not shown in Fig. 8(c) because it is much suppressed by the digital high-pass filter. It can also be seen that one cuboid and two cylinders can be clearly distinguished after the leakage cancellation. When the power of the echo signal is increased to -15.1 dBm, the difference between the echo signal power and the leakage signal power is reduced. The imaging results without and with leakage cancellation are shown in Fig. 8(e) and (f). When the power of the echo signal is further increased to -10.6 dBm, the corresponding results are shown in Fig. 8(g) and (h). Note that the influence of the background environment on the imaging results shown in Fig. 8(b) can also be observed in Fig.8 (c)-(h) due to the background environment. It can also be seen from Fig. 8 that, when the echo signal power is small, the imaging results after the leakage cancellation are still influenced by the residual leakage, the residual interference, and the background noise. When the echo signal power is increased, the influence of the residual leakage, the residual interference, and the background noise are decreased and more clear imaging results can be obtained after the leakage cancellation.

3.4 Leakage cancellation and distance measurement

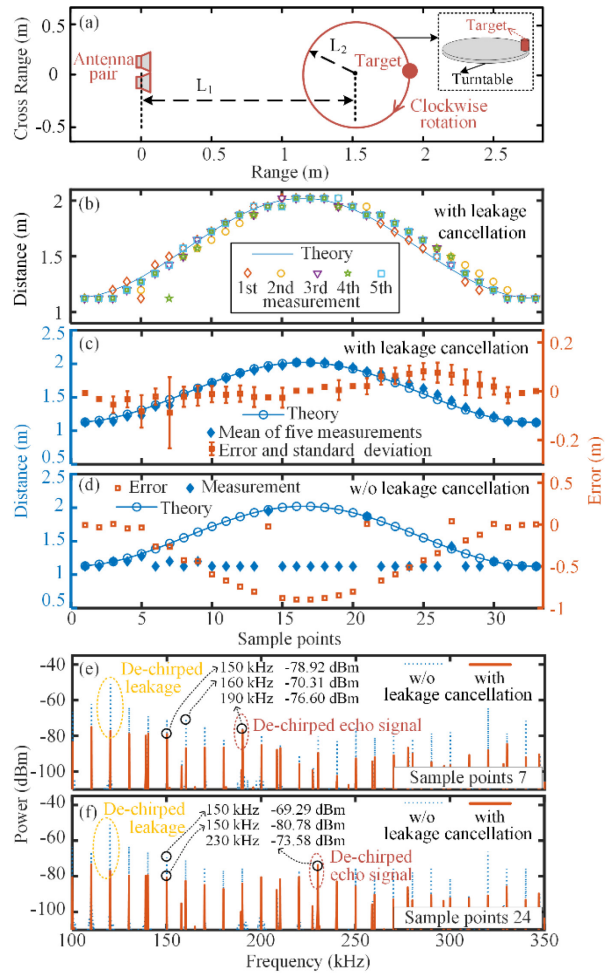

Fig. 9. (a) The schematic diagram of the distance measurement. (b) Five times distance measurement with leakage cancellation. (c) Measured distance and measurement errors with leakage cancellation. (d) Measured distance and measurement errors without leakage cancellation. Electrical spectra of the de-chirped signal with and without leakage cancellation in the (e) seventh and (f) twenty-fourth samples.

Finally, a signal target is placed on the turntable to verify the leakage cancellation performance through the target distance measurement. In this study, the center frequency and bandwidth of the LFM signal from the BPF are 11.5 and 2 GHz, the power of the leakage signal is -3 dBm, and the power of the de-chirp reference signal is changed to -12.8 dBm by tuning ATT1. One cylinder packed with silver paper is used as the target. The radius and height of the target are 5 and 15 cm, respectively. The target is placed on the turntable, as shown in Fig. 9(a), and the turntable rotates clockwise with a period of 24.56 s. During the rotation, the echo signal is sampled every 1/32 period with a 10 ms sampling time for a total of 33 samples. Five measurements are implemented with and without leakage cancellation. The distance $L_2$ between the cylinder and the center of the turntable is 45 cm, and the distance $L_1$ between the antenna pair and the center of the turntable is 155 cm. Since the range of movement of the target is around 110 cm to 200 cm, the distance measurement range is set from 109 cm to 225cm which corresponding de-chirped frequency range is from 145 kHz to 300 kHz. It should be noted that the frequency of the de-chirped leakage is 120 kHz.

The distance measurement results with leakage cancellation are shown in Fig. 9(b) and (c). Fig. 9(b) shows five times distance measurements. In Fig. 9(c), the mean distance values of the five measurements, the distance theoretical values, and the standard deviations of the distance measurement errors are represented by blue diamonds, blue circles, and the red error bars, respectively. Fig. 9(d) shows the results of one-time distance measurement without leakage cancellation. The measured value of the distances, the distance theoretical values, and measurement errors are represented by blue diamonds, blue circles, and red squares, respectively. It can be seen that the distance measurement can be achieved after leakage cancellation. In Fig. 9(c), the standard deviation of the fifth and seventh sampling is larger, because one or two corresponding measurements were incorrect due to the low echo signal power, as shown in Fig. 9(b). Excluding the three data that were measured incorrectly, the measurement error is about ±10 cm after leakage cancellation. As Fig. 9(d), when the leakage cancellation is not enabled, most of the measurements are wrong, it can be explained according to the result in Fig. 9(e) and Fig. 9(f). Fig. 9(e) and (f) show the electrical spectra of the de-chirped signal with and without leakage cancellation in the seventh and twenty-fourth samples. It can be seen that when the leakage is not canceled, the power at 150 or 160 kHz is always the maximum in the searched frequency range (from 145 kHz to 300 kHz). Thus, most of the measured distances are wrong when the leakage cancellation is not implemented. When the leakage is canceled, the noise is decreased and the frequency of the de-chirped echo signal can be correctly found, as shown by the red solid line in Fig. 9(e) and (f). To perform the leakage cancellation, the noise can be decreased and the radar measurement performance can be effectively improved.

## 4. Conclusion

In summary, we have demonstrated a leakage cancellation and echo signals de-chirping approach for FMCW radar systems. The key significance of the work is that, for the first time, it combines the leakage cancellation and echo signals de-chirping mainly in a single DD-MZM and the effect of leakage cancellation on the imaging and distance measurement of FMCW radar systems is studied. An experiment is performed. The cancellation performance of different de-chirp reference power and different echo signal power are studied, and the power of the de-chirp reference signal is adjusted to optimize the cancellation performance. The cancellation depth of the de-chirped signal is around 23 dB when the LFM center frequency and bandwidth are 11 and 2 GHz. After the leakage cancellation, the imaging results of multiple imaging targets can be clearly distinguished, and the error of the distance measurement results is significantly reduced to ±10 cm.

**Acknowledgements**

This work was supported by the National Natural Science Foundation of China [grant number 61971193]; the Natural Science Foundation of Shanghai [grant number 20ZR1416100]; the Open Fund of State Key Laboratory of Advanced Optical Communication Systems and Networks, Peking University, China [grant number 2020GZKF005]; the Science and Technology Commission of Shanghai Municipality [grant number 18DZ2270800].